\documentclass[eqsecnum,preprint,preprintnumbers,12pt]{revtex4}
\usepackage{subeqnarray,graphicx}

\begin{document}

\title{Analytic Form of the QCD Instanton Determinant for Small Quark Mass}
\author{Jin Hur} \email{hurjin@kias.re.kr}
\affiliation{School of Computational Sciences, Korea Institute for Advanced Study, Seoul 130-012, Korea}
\author{Choonkyu Lee} \email{cklee@phya.snu.ac.kr}
\affiliation{Department of Physics and Astronomy and Center for Theoretical Physics\\ Seoul National University, Seoul 151-742, Korea}
\author{Hyunsoo Min} \email{hsmin@dirac.uos.ac.kr}
\affiliation{Department of Physics, University of Seoul, Seoul 130-743, Korea}

\begin{abstract}
We use a novel method to calculate analytically the QCD instanton prefactor due to a quark field carrying a small mass parameter $m$. In the SU(2) instanton background of size $\rho$, the spinor effective action $\Gamma^F$ (in the minimal subtraction scheme), which gives rise to the prefactor $\exp (-\Gamma^F)$, is shown to have the small-$m \rho$ behavior
\begin{eqnarray}
\Gamma^F &=& -\ln (m /\mu) - \ln (\mu\rho)/3 -2 \,\alpha(1/2) -(m \rho)^2 \{ \ln (m \rho /2) +\gamma+1/2\}  \nonumber\\
&& -2 (m \rho)^4 \{ -\ln ^2(m \rho)/4+\ln (m \rho) (1/2-\gamma +\ln 2)/2+C \} +O ( (m \rho)^6 ), \nonumber
\end{eqnarray}
where $\gamma=0.577216\ldots$, $\alpha(1/2)=0.145873\ldots$, and our numerically evaluated value for the constant $C$ is $C=-0.382727\ldots$. A good agreement between this form and the numerically exact calculation is found if $(m \rho) \lesssim 0.8$.
\end{abstract}
\maketitle

\section{Introduction}
Fermion determinants in the background of certain nontrivial gauge fields play some important roles in connection with various quantum field-theoretic studies, but they are notoriously difficult to calculate analytically. At the heart of the problem lies the fact that this requires the knowledge on the spectral shifts for given partial differential operators with the introduction of background fields. But, for partial differential operators which are completely separable (as would be the case with generic radially-symmetric background fields), it has become possible to evaluate the corresponding functional determinants numerically with high precision. See for instance Refs. \cite{insdet,rea1,rea3} for a particularly efficient method developed recently by us (with G. Dunne). Our method, which is really a combination of analytic and numerical schemes, was first used to compute the QCD instanton determinant for an arbitrary value of quark mass \cite{insdet}. Then, applications to other cases involving more general classes of radial background fields were also made \cite{rea1,rea3,dunnemin}.

The single-instanton determinant due to a quark field of mass $m$ is essentially a function of $m \rho$ where $\rho$ is the instanton size (as the dependence on the renormalization mass scale $\mu$ can easily be separated). Aside from finding the precise numerical value for the determinant for given values of $m$ and $\rho$, it will then certainly be desirable, for practical physics applications for instance, to have an explicit (analytic) function form for the quantity in some limiting cases at least. For large $m \rho$ a systematic large mass expansion \cite{kwon} (accounted for most simply by using the heat-kernel expansion) provides such analytic formula, and for $m \rho \gtrsim 1.3$ this formula was shown to be a good representation of our numerically determined results. More interesting is the small $m \rho$-limit, as instanton effects are believed to be important for light quarks especially. 't Hooft, in a classic paper \cite{thooft}, first obtained the analytic expression appropriate to the $m \rho \to 0$ limit of the determinant and the result of \cite{insdet}, being essentially exact for all $m \rho$ values, is of course consistent with his expression. But, as we increase the value of $m \rho$ from zero, a small discrepancy was observed between the numerical result of \cite{insdet} and the formula based on the small-mass expansion of the determinant \cite{kwon,calitz} given in earlier literatures. As the latter expansion (utilizing the known exact massless particle propagator \cite{brown} in the instanton background) required rather delicate manipulation for certain integrals appearing due to the infrared problem, we had naturally our suspiction on this small-mass expansion. We thus decided to develop an entirely new approach for the small-mass behavior of the instanton determinant, which is the present work. Here we not only go beyond the previous finding but also introduce a new method which should have useful application in dealing with other related problems. Our new small-mass expansion formula is now fully consistent with the numerical result of \cite{insdet}.

In this paper we will discuss small-mass behaviors of the instanton determinant within the general strategy of Ref. \cite{insdet}, but now utilizing the (analytically developed) small-mass expansion for the very part reserved for numerical evaluation in Ref. \cite{insdet}, i.e., for low-partial-wave radial determinants. Note that these radial determinants are readily determined once one has the \textit{global} (i.e., valid over the entire radial range) solutions to the Gel'fand-Yaglom initial-value problems of Sturm-Liouville-type differential equations \cite{gelfand}. So the nontrivial part of our analysis concerns the development of globally-valid small-mass perturbation theory for the solution of the Gel'fand-Yaglom initial value problem. In view of the presence of two characteristic length scales, i.e., $\rho$ and $\frac{1}{m}$, we here find such globally-valid approximation by combining one perturbative solution valid in the radial range $0<r \lesssim R$ (here, $R$ can be any value satisfying the condition $\rho \ll R \ll \frac{1}{m}$) with another perturbative development valid in the asymptotic range $r \gtrsim R$, through an appropriate \textit{matching} procedure. (A similar approach has been employed to discuss the 't Hooft-Polyakov magnetic monopole solution for small, but nonzero, Higgs mass in Ref \cite{gardner}).

This paper is organized as follows. In Sec. \ref{secreview} we give a brief review on the calculation scheme used in \cite{insdet,rea1,rea3} and also collect useful results for our later discussions. Then, in Sec. \ref{secperturbation}, we present a detailed discussion on how the desired small-mass perturbative solution to the Gel'fand-Yaglom initial value problem can be secured by combining perturbative developments derived in distinct radial regions (through matching them in the region of common validity). These results are then used to determine small-mass behaviors of the QCD instanton determinant explicitly up to $O((m \rho)^4)$ in Sec. \ref{secsmall}. In Sec. \ref{secdiscussion} we discuss our findings against the numerical result of Ref. \cite{insdet} and conclude with some relevant remarks.

\section{Review of some relevant results} \label{secreview}
Because of a hidden supersymmetry \cite{thooft}, the spinor Dirac operator in an SU(2) single instanton field \cite{belavin}
\begin{eqnarray}
A_\mu (x) = \frac{\eta_{\mu\nu a} \tau^a x_\nu}{r^2+\rho^2}, \qquad (r \equiv \sqrt{x_\mu x_\mu}) \label{instantonsolution}
\end{eqnarray}
has the same spectrum (except for zero modes and an overall multiplicity factor of 4) as the corresponding scalar Klein-Gordon operator. Therefore the one-loop effective action, i.e., minus the logarithm of the functional determinant for a Dirac spinor field of mass $m$ (and isospin $\frac{1}{2}$), $\Gamma^F (A;m)$, is directly related to the corresponding scalar effective action $\Gamma^S (A;m)$ (for a complex scalar of mass $m$ and isospin $\frac{1}{2}$) by \cite{kwon}
\begin{eqnarray}
\Gamma^F (A;m) = - \ln \frac{m}{\mu} -2 \Gamma^S (A;m), \label{gammaF}
\end{eqnarray}
where $\mu$ is the renormalization scale. The renormalization prescription for the effective actions here is that of minimal subtraction. Thus, for any given mass value $m$, it suffices to consider the scalar effective action to learn about the corresponding fermion effective action. Also useful is the fact that $\Gamma^S (A;m)$ can be related to another quantity $\tilde{\Gamma}^S (m \rho)$, which is a function of $m \rho$ only, by
\begin{eqnarray}
\Gamma^S (A;m) = \frac{1}{6} \ln (\mu \rho) +\tilde{\Gamma}^S (m \rho). \label{gammaS}
\end{eqnarray}
(The factor $\frac{1}{6}$ in (\ref{gammaS}) is determined by the one-loop $\beta$-function). Then, concentrating on the $m \rho$ dependence of $\tilde{\Gamma}^S (m \rho)$, we can set the instanton size to unity, i.e., $\rho=1$.

To determine the effective action $\Gamma^S (A;m)$ in the single instanton background (\ref{instantonsolution}), one should be able to evaluate individual partial wave contributions
\begin{eqnarray}
\Gamma_J (m) \equiv \ln \!\left( \frac{\det (\mathcal{H}_J +m^2)}{\det (\mathcal{H}^{\text{free}}_l +m^2)} \right). \label{partialeffectiveaction}
\end{eqnarray}
Here, $J=(l,j)$ ($l=0,\frac{1}{2},1,\frac{3}{2},\cdots$; $j=\left| l \pm \frac{1}{2} \right|$) represent quantum numbers needed to designate given partial waves, and $\mathcal{H}_{(l,j)}$, $\mathcal{H}^{\text{free}}_l$ are radial differential operators given by
\begin{eqnarray}
\mathcal{H}_{(l,j)} &=& -\frac{\partial^2}{\partial r^2} +\frac{4l(l+1)+\frac{3}{4}}{r^2} +\frac{4(j-l)(j+l+1)}{r^2+1} -\frac{3}{(r^2+1)^2} \nonumber\\
&\equiv& -\frac{\partial^2}{\partial r^2} +V_{\text{eff}} (r), \label{Hlj} \\
\mathcal{H}_l^{\text{free}} &=& -\frac{\partial^2}{\partial r^2} +\frac{4l(l+1)+\frac{3}{4}}{r^2},
\end{eqnarray}
respectively. (Our notations here slightly differ from those of Ref. \cite{insdet}; $\mathcal{H}_{(l,j)}$ and $\mathcal{H}^{\text{free}}_l$ of the present paper do not represent the covariant Laplacian $-D^2$ in the given partial wave (as in Ref. \cite{insdet}), but the corresponding differential operators after extracting a measure factor $r^{3/2}$, i.e., coincide with the operators $\tilde{\mathcal{H}}_{(l,j)}$ and $\tilde{\mathcal{H}}^{\text{free}}_l$ of Ref. \cite{rea1}). Such knowledge on individual partial wave contributions may then be used to determine $\tilde{\Gamma}^S (m)$ with the help of the formula (first derived in Ref. \cite{insdet})
\begin{eqnarray}
\tilde{\Gamma}^S (m)&=&\lim_{L\to \infty } \left[\sum _{l=0,\frac{1}{2},\cdots }^L (2 l+1) (2 l+2) \left\{\Gamma_{(l,j=l+\frac{1}{2})}(m)+\Gamma_{(l+\frac{1}{2},j=l)}(m) \right\} +\Gamma _{l>L}(m)\right], \quad \label{effectiveaction}
\end{eqnarray}
where the partial wave contribution to the effective action, $\Gamma_{(l,j)}(m)$, is defined by (\ref{partialeffectiveaction}). The large partial wave contribution (with the renormalization counter-term included), $\Gamma _{l>L}(m)$, was calculated to have the following explicit form:
\begin{eqnarray}
\Gamma _{l>L}(m)=2 L^2+4 L-\left(\frac{1}{6}+\frac{m^2}{2}\right) \ln L+\frac{127}{72}-\frac{\ln 2}{3}+\frac{m^2}{2} \left( \ln \frac{m}{4} +1 \right). \label{largeL}
\end{eqnarray}
To achieve faster convergence for $L \to \infty$ in (\ref{effectiveaction}), one can also include higher WKB terms in $\Gamma_{l>L}(m)$ as discussed in \cite{rea1,rea3}; but, we have no immediate use for such improvement terms in this work.

To evaluate the partial wave contribution $\Gamma_{(l,j)} (m)$ for $l \leq L$, the method due to Gel'fand and Yaglom \cite{gelfand} can be used. That is, given two radial differential operators $\mathcal{M}_1=\mathcal{H}_{(l,j)}+m^2$ and $\mathcal{M}_2=\mathcal{H}_l^{\text{free}}+m^2$ on the interval $r \in [0,\infty)$, the ratio of their determinants is given by
\begin{eqnarray}
\frac{\det \mathcal{M}_1}{\det \mathcal{M}_2} = \lim_{\mathbf{R} \to \infty} \left[ \frac{\phi(\mathbf{R})}{\phi^{\text{free}}(\mathbf{R})} \right], \label{GYtheorem}
\end{eqnarray}
where $\phi(r),\phi^{\text{free}}(r)$ satisfy the initial value problems:
\begin{eqnarray}
\left(\mathcal{H}_{(l,j)}+m^2\right) \phi (r) = 0, \qquad \phi (r)\to r^{2 l+\frac{3}{2}} \text{ as } r\to 0 \label{GYequation}
\end{eqnarray}
and
\begin{eqnarray}
\left(\mathcal{H}^{\text{free}}_{l}+m^2\right) \phi^{\text{free}} (r) = 0, \qquad \phi^{\text{free}} (r)\to r^{2 l+\frac{3}{2}} \text{ as } r\to 0. \label{GYequationfree}
\end{eqnarray}
With nonvanishing mass $m$ the analytic solution to (\ref{GYequation}) is not known, and hence in Ref. \cite{insdet} an extensive numerical analysis was necessary to find $\phi(\mathbf{R})$ when $\mathbf{R}$ is large. With $m=0$, however, the explicit solution to the given initial value problem for general $l,j$ is found in terms of the hypergeometric function
\begin{eqnarray}
\phi_0 (r) \equiv [\phi(r)]_{m=0}=\frac{r^{2 l+\frac{3}{2}}}{\sqrt{r^2+1}} \,_2 F_1 \!\left(-j+l-\frac{1}{2}, j+l+\frac{1}{2};2 l+2;-r^2\right),
\end{eqnarray}
and consequently we have
\begin{eqnarray}
\phi_0^+(r) &\equiv& [\phi_0(r)]_{(l,j=l+\frac{1}{2})}=\frac{r^{2 l+\frac{3}{2}} \left\{(2 l+1) r^2+2 l+2\right\}}{2 (l+1) \sqrt{r^2+1}}, \label{pluszerosol} \\
\phi_0^-(r) &\equiv& [\phi_0(r)]_{(l+\frac{1}{2},j=l)}=\frac{r^{2 l+\frac{5}{2}}}{\sqrt{r^2+1}}. \label{minuszerosol}
\end{eqnarray}
On the other hand, the solution to (\ref{GYequationfree}) for any mass value $m$ is
\begin{eqnarray}
\phi^{\text{free}} (r)=\frac{(2 l+1)!}{\sqrt{2}} \left(\frac{2}{m}\right)^{2 l+\frac{3}{2}} \sqrt{m r}\; I_{2 l+1}(m r), \label{freesolution}
\end{eqnarray}
and hence, with $m=0$, the expression
\begin{eqnarray}
\phi^{\text{free}}_0(r) = r^{2 l+\frac{3}{2}}. \label{freezerosolution}
\end{eqnarray}
If one uses the results provided by (\ref{pluszerosol}), (\ref{minuszerosol}) and (\ref{freezerosolution}) with our formulas (\ref{GYtheorem}) and (\ref{effectiveaction}), the massless limit form of 't Hooft for the instanton determinant can be obtained.

\section{Mass perturbation with the Gel'fand-Yaglom initial value problem} \label{secperturbation}
In this section we will find the asymptotic behavior of the solution to (\ref{GYequation}) when $m$ is small. To this end, the small-$m$ power series development for $\phi(r)$ may be considered, with the exact $m=0$ solution given in (\ref{pluszerosol}) or (\ref{minuszerosol}) as the zeroth order solution. A naive construction of such series works fine as long as $r$ is taken to be finite, i.e., is applicable in the range $0<r \lesssim R$ where $\rho(=1) \ll R \ll \frac{1}{m}$. We will refer to this series solution as the small-$r$ solution below. The small-$r$ solution breaks down if $r$ becomes comparable to $\frac{1}{m}$, and hence it cannot be used to find the asymptotic form $\phi(\mathbf{R})$ (needed to apply (\ref{GYtheorem})). To overcome this, we will introduce an alternative perturbation series in $m$, which we will call as the large-$r$ solution. The range of application for the latter series is for any $r \gtrsim R$ with $R \gg 1$. We will then connect this large-$r$ solution to our small-$r$ solution in the intermediate region (represented by $r=R$) i.e., match the small-$r$ and large-$r$ solutions to obtain the globally valid solution to our initial value problem. The asymptotic form $\phi(\mathbf{R})$ can be determined from this global solution. Finding a global solution in this way of matching is known as the boundary layer method. For generic mathematical discussions about this method, see \cite{bender}.

First we describe a general method to construct the small-$m$ perturbative solution to the initial value problem associated with an differential equation
\begin{eqnarray}
\left(-\frac{\partial^2}{\partial r^2}+V(r)\right) f (r) = 0, \label{sampleq}
\end{eqnarray}
where $V(r)$ can have a series form
\begin{eqnarray}
V(r) = V_0(r) +m^2 V_1(r) +m^4 V_2(r) +\cdots \;. \label{Vser}
\end{eqnarray}
We here assume that the corresponding solution in the $m=0$ limit is known explicitly. We write the perturbative solution as
\begin{eqnarray}
f (r)=f_0(r)+m^2 f_1(r)+m^4 f_2(r)+\cdots \;, \label{series}
\end{eqnarray}
and then the leading term in the series, $f_0(r) \equiv [f(r)]_{m=0}$, satisfies the differential equation
\begin{eqnarray}
\left(-\frac{\partial^2}{\partial r^2}+V_0(r)\right) f_0 (r) = 0. \label{sampleq0}
\end{eqnarray}
Using the equations of motion (\ref{sampleq}) and (\ref{sampleq0}), we now have the relation
\begin{eqnarray}
\frac{\partial }{\partial r}\left\{f'(r) f_0(r)-f(r) f_0'(r)\right\} &=& f''(r) f_0(r)-f(r) f_0''(r) \nonumber\\
&=& \left\{ V(r)-V_0(r) \right\} f (r) f_0(r). \label{identity}
\end{eqnarray}
Plugging in the series (\ref{series}) into (\ref{identity}) then yields the recurrence relations
\begin{eqnarray}
\frac{\partial }{\partial r}\left\{f_k'(r) f_0(r)-f_k(r) f_0'(r)\right\}= \sum_{i=1}^k V_i(r) f_{k-i}(r) f_0(r), \qquad (k=1,2,\cdots\;). \label{recurrence}
\end{eqnarray}
Integrating (\ref{recurrence}) from a certain `initial' point $r^*$ to $r$, we find
\begin{eqnarray}
&& f_k'(r) f_0(r)-f_k(r) f_0'(r) \nonumber\\
&& = f_k'(r^*) f_0(r^*)-f_k(r^*) f_0'(r^*) +\int _{r^*}^rdt \sum_{i=1}^k V_i(t) f_{k-i}(t) f_0(t). \label{firstintegral}
\end{eqnarray}
Since the left hand side of (\ref{firstintegral}) is equal to $f_0 (r)^2 \frac{\partial}{\partial r} [f_k (r)/f_0(r)]$, we can integrate this equation once more to obtain the following expression for $f_k(r)$:
\begin{eqnarray}
f_k(r) &=& f_0(r) \left[ \frac{f_k(r^*)}{f_0(r^*)} + \int_{r^*}^r \frac{du}{f_0(u){}^2} \Bigg\{ f_k'(r^*) f_0(r^*)-f_k(r^*) f_0'(r^*) \right. \nonumber\\
&& \qquad\qquad\qquad\qquad \left.\left. +\int _{r^*}^u dt \sum_{i=1}^k V_i(t) f_{k-i}(t) f_0(t) \right\} \right], \qquad (k=1,2,\cdots\;). \label{recurrenceintegral}
\end{eqnarray}
These equations may be used successively to find $f_1(r)$, $f_2(r)$, $\cdots$ for $r\geq r^*$ if the function $f_0(r)$ and `initial' values $f_1(r^*)$, $f_1'(r^*)$, $f_2(r^*)$, $f_2'(r^*)$, $\cdots$ are given. But the uniform convergence of the above series (\ref{series}) in a specified domain of $r$ is an issue that should be settled separately.

\subsection{The small-$r$ solution}
Now consider our equation (\ref{GYequation}), i.e.,
\begin{eqnarray}
\left(-\frac{\partial ^2}{\partial r^2}+V_{\text{eff}}(r)+m^2\right) \phi (r) &=& 0, \qquad \phi (r)\to r^{2 l+\frac{3}{2}} \text{ as } r\to 0 \label{phieq}
\end{eqnarray}
with $V_{\text{eff}}(r)$ defined in (\ref{Hlj}). To find the appropriate series solution
\begin{eqnarray}
\phi (r)=\phi_0(r)+m^2 \phi_1(r)+m^4 \phi_2(r)+\cdots \label{phiser}
\end{eqnarray}
by the above method, it is to be noted that we have here $V_0(r) = V_{\text{eff}}(r)$, $V_1(r) = 1$, and $V_k(r)=0$ if $k\geq 2$, in the power series notation of (\ref{Vser}). Also, since the initial condition is given at the origin, we take $r^* \to 0$ and may impose, from (\ref{GYequation}),
\begin{eqnarray}
r^* \to 0 \;:\qquad \frac{\phi_k(r^*)}{\phi_0(r^*)} \to 0, \qquad \text{for } k=1,2,\cdots \;.
\end{eqnarray}
With this information we can use (\ref{recurrenceintegral}) to find $\phi_1(r)$, $\phi_2(r)$, $\cdots$ successively. For the first-order term, i.e., for
\begin{eqnarray}
\phi _1(r)=\phi _0(r) \int _0^r\frac{du}{\phi _0(u){}^2} \int _0^udt \; \phi _0(t){}^2 \label{phifirstintegral}
\end{eqnarray}
with $\phi_0(r)$ given in (\ref{pluszerosol}) or (\ref{minuszerosol}), the integrals can be performed explicitly, to obtain the expressions
\begin{eqnarray}
&& \phi_1^+(r)=\phi _0^+(r) \left[\frac{r^2}{8 (l+1)}+\frac{\ln \!\left(r^2+1\right)}{8 (l+1) (2 l+1)}-\frac{r^2}{(2 l+1) r^2+2 l+2} \right. \nonumber\\
&& \qquad \times \left\{\frac{1}{4 (2 l+1)} \,_2 F_1 \!\left(1,2 l+4;2 l+5;-r^2\right)+\frac{r^2}{2 (2 l+3)} \,_2 F_1 \!\left(1,2 l+3;2 l+4;-r^2\right) \right. \nonumber\\
&& \qquad\qquad \left.\left. +\frac{(2 l+1) r^4}{16 (l+1) (l+2)} \,_2 F_1 \!\left(1,2 l+4;2 l+5;-r^2\right)\right\}\right], \label{phi1plussol} \\
&& \phi _1^-(r)=\phi _0^-(r) \left[\frac{r^2}{4 (2 l+1)}-\frac{\ln \!\left(r^2+1\right)}{8 (l+1) (2 l+1)} \right. \nonumber\\
&& \qquad\qquad \left. -\frac{r^2 \left\{(2 l+2) r^2+2 l+1\right\}}{8 (l+1) (2 l+1) (2 l+3)} \,_2 F_1 \!\left(1,2 l+3;2 l+4;-r^2\right) \right], \label{phi1minussol}
\end{eqnarray}
where we have denoted $[\phi_1(r)]_{J=(l,j=l+1/2)}$ and $[\phi_1(r)]_{J=(l+1/2,j=l)}$ by $\phi_1^+(r)$ and $\phi_1^-(r)$, respectively (in accordance with the convention used in (\ref{pluszerosol}) and (\ref{minuszerosol}) for $\phi_0(r)$). Finding the next order term, i.e.,
\begin{eqnarray}
\phi _2(r)=\phi _0(r) \int _0^r\frac{du}{\phi _0(u){}^2} \int _0^u dt\; \phi _0(t) \phi _1(t), \label{phi2integral}
\end{eqnarray}
is not easy. However, by using the asymptotic series for $\phi_0(r)$ and $\phi_1(r)$ in (\ref{phi2integral}), we can at least obtain the asymptotic (i.e., valid for sufficiently large-$r$) series development for $\phi_2(r)$ with some effort. We will describe this calculation below.

First, from (\ref{pluszerosol}), (\ref{minuszerosol}), (\ref{phi1plussol}) and (\ref{phi1minussol}), the following large-$r$ asymptotic series for $\phi_0(r)$ and $\phi_1(r)$ can readily be deduced:
\begin{eqnarray}
\phi _0^+(r)&=&r^{2 l+\frac{5}{2}} \left\{\frac{2 l+1}{2 l+2}+\frac{2 l+3}{(4 l+4) r^2}-\frac{2 l+5}{16 (l+1) r^4}+\cdots \right\}, \label{phi0plusser} \\
\phi _0^-(r)&=&r^{2 l+\frac{3}{2}} \left\{1-\frac{1}{2 r^2}+\frac{3}{8 r^4}+\cdots \right\}, \\
\phi _1^+(r)&=&r^{2 l+\frac{9}{2}} \left\{\frac{2 l+1}{8 (l+1) (2 l+3)}+\frac{1}{r^2} \left(\frac{\ln r}{8 (l+1)^2}+\frac{4 l^3+14 l^2+12 l+1}{32 (l+1)^3 (2 l+3)}\right) \right. \nonumber\\
&& \left. +\frac{1}{r^4} \left(\frac{(2 l+3) \ln r}{16 (l+1)^2 (2 l+1)}-\frac{l \left(8 l^4+12 l^3-42 l^2-87 l-40\right)}{64 (l+1)^3 (2 l+1)^2 (2 l+3)}\right)+\cdots \right\}, \\
\phi _1^-(r)&=&r^{2 l+\frac{7}{2}} \left\{\frac{1}{8 (l+1)}+\frac{1}{r^2} \left(-\frac{\ln r}{4 (l+1) (2 l+1)}-\frac{4 l^3+8 l^2+l-2}{16 (l+1)^2 (2 l+1)^2}\right) \right. \nonumber\\
&& \left. +\frac{1}{r^4} \left(\frac{\ln r}{8 (l+1) (2 l+1)}+\frac{12 l^4+8 l^3-33 l^2-35 l-8}{64 l (l+1)^2 (2 l+1)^2}\right)+\cdots \right\}. \label{phi1minusser}
\end{eqnarray}
If these series are used with the function inside the $u$-integral (\ref{phi2integral}), it is not difficult to see that its large-$r$ asymptotic behavior (for given $J=(l,j=l+\frac{1}{2})$ or $J=(l+\frac{1}{2},j=l)$) is described by the form
\begin{eqnarray}
\frac{1}{\phi _0(r){}^2} \int _0^rdt\; \phi _0(t) \phi _1(t)=A_1 r^3+\left(B_2 \ln r+A_2\right) r+\frac{B_3 \ln r+A_3}{r}+\cdots \;, \label{firstseriesintegral}
\end{eqnarray}
where $A_k$'s and $B_k$'s are some constants. By integrating this series, we may then conclude that
\begin{eqnarray}
&& \int_0^r \frac{du}{\phi_0(u){}^2} \int _0^udt\; \phi_0(t) \phi_1(t) \nonumber\\
&& \quad =\frac{A_1}{4} r^4+\left(\frac{B_2}{2} \ln r-\frac{B_2}{4}+\frac{A_2}{2}\right) r^2+\frac{B_3}{2} \ln ^2 r+A_3 \ln r+C_1+\cdots \;, \label{secondseriesintegral}
\end{eqnarray}
where $C_1$ is a certain constant. The constant $C_1$ is not given by this consideration which relies just on the asymptotic forms of $\phi_0(r)$ and $\phi_1(r)$; but, one can evaluate it based on the representation
\begin{eqnarray}
C_1&=&\lim_{r\to \infty } \left[\int _0^r\frac{du}{\phi_0(u){}^2} \int _0^udt\; \phi_0(t) \phi_1(t) \right. \nonumber\\
&& \qquad \left. -\left\{\frac{A_1}{4} r^4+\left(\frac{B_2}{2} \ln r-\frac{B_2}{4}+\frac{A_2}{2}\right) r^2+\frac{B_3}{2} \ln ^2 r+A_3 \ln r\right\}\right], \label{C1def}
\end{eqnarray}
as described in the Appendix. By using the large-$r$ form (\ref{secondseriesintegral}) with (\ref{recurrenceintegral}), the large-$r$ asymptotic behaviors for $\phi_2(r)$ can be found. Explicitly, for sufficiently large-$r$, we have the series
\begin{eqnarray}
\phi _2^+(r)&=&r^{2 l+\frac{13}{2}} \left\{\frac{2 l+1}{128 (l+1) (l+2) (2 l+3)} \right. \nonumber\\
&& \quad +\frac{1}{r^2} \left(\frac{\ln r}{32 (l+1)^2 (2 l+3)} +\frac{4 l^4+20 l^3+17 l^2-28 l-31}{256 (l+1)^3 (l+2) (2 l+3)^2}\right) \nonumber\\
&& \quad +\frac{1}{r^4} \left(\frac{\ln^2 r}{64 (l+1)^3 (2 l+1)} +\frac{\left(4 l^4+28 l^3+43 l^2+19 l+1\right) \ln r}{64 (l+1)^4 (2 l+1)^2 (2 l+3)} \right. \nonumber\\
&& \qquad \left.\left. -\frac{4 l^4+20 l^3+73 l^2+138 l+85}{1024 (l+1)^3 (l+2) (2 l+1) (2 l+3)}+\frac{2 l+1}{2 l+2} C_1^+(l)\right)+\cdots \right\}, \qquad \label{phi2plusser}\\
\phi _2^-(r)&=&r^{2 l+\frac{11}{2}} \left\{\frac{1}{64 (l+1) (2 l+3)} \right. \nonumber\\
&& \quad +\frac{1}{r^2} \left(-\frac{\ln r}{32 (l+1)^2 (2 l+1)}-\frac{4 l^4+12 l^3-7 l^2-32 l-15}{128 (l+1)^3 (2 l+1)^2 (2 l+3)}\right) \nonumber\\
&& \quad +\frac{1}{r^4} \left(\frac{\ln^2 r}{32 (l+1)^2 (2 l+1)^2}+\frac{\left(4 l^4-4 l^3-29 l^2-23 l-4\right) \ln r}{64 l (l+1)^3 (2 l+1)^3} \right. \nonumber\\
&& \qquad \left.\left. +\frac{12 l^4+36 l^3-l^2-58 l-29}{512 (l+1)^3 (2 l+1)^2 (2 l+3)}+C_1^-(l)\right)+\cdots \right\}, \qquad (l\neq 0) \label{phi2minusser}
\end{eqnarray}
and
\begin{eqnarray}
\left[\phi _2^-(r)\right]{}_{l=0}&=&r^{\frac{11}{2}} \left\{\frac{1}{192}+\frac{1}{r^2} \left(-\frac{\ln r}{32}+\frac{5}{128}\right) \right. \nonumber\\
&& \quad \left. +\frac{1}{r^4} \left(-\frac{3 \ln^2 r}{32}+\frac{5 \ln r}{64}-\frac{29}{1536}+C_1^-(l=0)\right)+\cdots \right\}. \label{phi2minus0ser}
\end{eqnarray}
In (\ref{phi2plusser}) and (\ref{phi2minusser}), $C_1^\pm(l)$ denote the constants which can be found with the help of the relation like (\ref{C1def}).

The above asymptotic behaviors of $\phi_2^\pm(r)$ may be compared with the asymptotics of $\phi_0^\pm(r)$ and $\phi_1^\pm(r)$, given in (\ref{phi0plusser})-(\ref{phi1minusser}). Observe that, as $r$ becomes large, $\phi_k(r)$ shows a power increase in $r$ (with the leading power increasing with $k$). Therefore, if $r$ becomes comparable to $\frac{1}{m}$, the series (\ref{phiser}) will not be applicable. Thus we may utilize the series expansion (\ref{phiser}) for $r$ in the range $0<r \lesssim R$ where $R$ can be any value satisfying the condition $1(=\rho) \ll R \ll \frac{1}{m}$.

\subsection{The large-$r$ solution and matching}
For $r \geq R$, we may contemplate on another approximation scheme. First, since $r \gg 1$, we can here approximate the potential $V_{\text{eff}}(r)$, defined in (\ref{Hlj}), by
\begin{eqnarray}
V_{\text{eff}}(r)=\frac{4 j(j+1)+\frac{3}{4}}{r^2}+\frac{-4 j (j+1)+4 l (l+1)-3}{r^4}+\cdots \;.
\end{eqnarray}
If we consider only the leading term of this potential, our equation (\ref{phieq}) reduces to the massive free equation (\ref{GYequationfree}) with $j$ instead of $l$. Then, motivated by the form (\ref{freesolution}) for the free solution, we introduce the rescaled independent variable $x=m r$ and consider a function $\varphi (x)\equiv \phi (r=\frac{x}{m})$ to recast (\ref{phieq}) as
\begin{eqnarray}
\left\{-\frac{\partial ^2}{\partial x^2}+1+\frac{1}{m^2} V_{\text{eff}}\left(\frac{x}{m}\right)\right\} \varphi (x)=0. \label{varphieq}
\end{eqnarray}
Now note that, for $m\to 0$, we have the expansion
\begin{eqnarray}
1+ \frac{1}{m^2} V_{\text{eff}}\left(\frac{x}{m}\right) &=& \left[1+\frac{4j(j+1)+\frac{3}{4}}{x^2} \right] +m^2 \left[ \frac{-4 j (j+1)+4 l (l+1)-3}{x^4} \right] \nonumber\\
&& +m^4 \left[ \frac{4 j (j+1)-4 l (l+1)+6}{x^6} \right]+\cdots \;, \label{vxser}
\end{eqnarray}
i.e., the equation (\ref{varphieq}) is of the very form for which our general perturbation method, described in (\ref{sampleq})-(\ref{recurrenceintegral}), can be applied. Therefore, representing the solution to (\ref{varphieq}) by the series
\begin{eqnarray}
\varphi (x)=\varphi_0(x)+m^2 \varphi_1(x)+m^4 \varphi_2(x)+\cdots \;, \label{varphiser}
\end{eqnarray}
we may use our iterative relations (\ref{recurrenceintegral}) (with the replacements $r\to x$, $r^* \to x^*$) to determine the functions $\varphi_1(x), \varphi_2(x), \cdots$ from the given zeroth order function $\varphi_0(x)$ and appropriate initial values given at $x=x^*$. For this it should be understood that
\begin{eqnarray}
V_0(r) &\to& V_0(x) = 1+\frac{4 j (j+1)+\frac{3}{4}}{x^2}, \\
V_1(r) &\to& V_1(x) = \frac{-4 j (j+1)+4 l (l+1)-3}{x^4}, \\
V_2(r) &\to& V_2(x) = \frac{4 j (j+1)-4 l (l+1)+6}{x^6}, \quad \cdots \;.
\end{eqnarray}

What value should we choose for the initial point $x^*$? We here remark that the small-$m$ expansion used in (\ref{vxser}) is evidently equivalent to a large-$x$ expansion of $V_{\text{eff}}(\frac{x}{m})$, and the series development (\ref{varphiser}) would become completely dubious for extremely small $x$. But, for $x$ moderately small so that we still have $x \gg m$ (i.e., for $r=R \gg 1$), there is no reason to suspect the validity of the series expansion (\ref{varphiser}). We thus choose our initial point at $x^* = m R$ (for $R$ satisfying condition $1 \ll R \ll \frac{1}{m}$), and determine $\varphi(x)$ for any $x \gtrsim m R$ ($x$ can be larger than $\frac{1}{m}$) by using the series (\ref{varphiser}) the explicit form of which is found iteratively using the relation (\ref{recurrenceintegral}). We can use the function $\phi(r) \equiv \varphi(x=m r)$ to obtain the large-$r$ solution to (\ref{GYequation}). For the zeroth order solution, we here take the (regular) free solution
\begin{eqnarray}
\varphi _0(x)= (\text{const.}) \frac{(2 j+1)!}{\sqrt{2}} \left(\frac{2}{m}\right)^{2 j+\frac{3}{2}} \sqrt{x}\; I_{2 j+1}(x), \label{varphi0}
\end{eqnarray}
To be general, one may allow $\varphi_0(x)$ to contain also the singular free solution (involving the Bessel function of the second kind, $K_{2 j+1}(x)$). But, in our case, such, more general, solution cannot be matched smoothly to our small-$r$ solution, which does not exhibit singularity for finite $r$, and by this reason it is sufficient to assume the form in (\ref{varphi0}).

To determine precisely the functions $\varphi_1(x)$, $\varphi_2(x)$, $\cdots$ by using (\ref{recurrenceintegral}), we need to fix the overall constant (in (\ref{varphi0})) and also specify the initial values $\varphi_1(m R)$, $\varphi_1'(m R)$, $\varphi_2(m R)$, $\varphi_2'(m R)$, $\cdots$. This is where we must demand on the solution $\varphi(x)$ a smooth matching to our small-$r$ solution $\phi(r)$ at $r=R$ (or, equivalently, at $x=m R$). That is, we require that, for $R$ satisfying the condition $1 \ll R \ll \frac{1}{m}$,
\begin{eqnarray}
&& \varphi _0(m R)+m^2 \varphi _1(m R)+m^4 \varphi _2(m R)+\cdots \nonumber\\
&& =\phi _0(R)+m^2 \phi _1(R)+m^4 \phi _2(R)+\cdots \;, \label{match}
\end{eqnarray}
where $\phi_0(R)$, $\phi_1(R)$ and $\phi_2(R)$ are given by the expressions in (\ref{phi0plusser})-(\ref{phi1minusser}), (\ref{phi2plusser})-(\ref{phi2minus0ser}) with $r=R$. (Note that, our $R$ denoting a generic value satisfying the condition $1 \ll R \ll \frac{1}{m}$, (\ref{match}) may be viewed as an equation for the `variable' $R$). We remark that there is a crucial consistency check (to be discussed below) on our matching condition (\ref{match}). Based on the condition (\ref{match}), the initial values can then be readily identified:
\begin{eqnarray}
&& \varphi _0^+(m R)=R^{2 l+\frac{5}{2}} \left\{\frac{2 l+1}{2 l+2}+\frac{(2 l+1) m^2 R^2}{8 (l+1) (2 l+3)}+\frac{(2 l+1) m^4 R^4}{128 (l+1) (l+2) (2 l+3)}+\cdots \right\}, \qquad\quad \label{varphi0ip} \\
&& \varphi _0^-(m R)=R^{2 l+\frac{3}{2}} \left\{1+\frac{m^2 R^2}{8 (l+1)}+\frac{m^4 R^4}{64 (l+1) (2 l+3)}+\cdots \right\}, \label{varphi0im} \\
&& \varphi _1^+(m R)=R^{2 l+\frac{5}{2}} \left\{\frac{2 l+3}{4 (l+1) m^2 R^2}+\left(\frac{\ln R}{8 (l+1)^2}+\frac{4 l^3+14 l^2+12 l+1}{32 (l+1)^3 (2 l+3)}\right) \right. \nonumber\\
&& \qquad\qquad \left. +m^2 R^2 \left(\frac{\ln R}{32 (l+1)^2 (2 l+3)}+\frac{4 l^4+20 l^3+17 l^2-28 l-31}{256 (l+1)^3 (l+2) (2 l+3)^2}\right)+\cdots \right\}, \label{varphi1ip} \\
&& \varphi _1^-(m R)=R^{2 l+\frac{3}{2}} \left\{-\frac{1}{2 m^2 R^2}+\left(-\frac{\ln R}{4 (l+1) (2 l+1)}-\frac{4 l^3+8 l^2+l-2}{16 (l+1)^2 (2 l+1)^2}\right) \right. \nonumber\\
&& \qquad\qquad \left. +m^2 R^2 \left(-\frac{\ln R}{32 (l+1)^2 (2 l+1)}-\frac{4 l^4+12 l^3-7 l^2-32 l-15}{128 (l+1)^3 (2 l+1)^2 (2 l+3)}\right)+\cdots \right\}, \label{varphi1im} \\
&& \varphi _2^+(m R)=R^{2 l+\frac{5}{2}} \left\{-\frac{2 l+5}{16 (l+1) m^4 R^4} \right. \nonumber\\
&& \qquad\qquad +\frac{1}{m^2 R^2} \left(\frac{(2 l+3) \ln R}{16 (l+1)^2 (2 l+1)}-\frac{l \left(8 l^4+12 l^3-42 l^2-87 l-40\right)}{64 (l+1)^3 (2 l+1)^2 (2 l+3)}\right) \nonumber\\
&& \qquad\qquad +\left(\frac{\ln ^2R}{64 (l+1)^3 (2 l+1)}+\frac{\left(4 l^4+28 l^3+43 l^2+19 l+1\right) \ln R}{64 (l+1)^4 (2 l+1)^2 (2 l+3)} \right. \nonumber\\
&& \qquad\qquad\qquad \left.\left. -\frac{4 l^4+20 l^3+73 l^2+138 l+85}{1024 (l+1)^3 (l+2) (2 l+1) (2 l+3)}+\frac{2 l+1}{2 l+2} C_1^+(l) \right)+\cdots \right\}, \\
&& \varphi _2^-(m R)=R^{2 l+\frac{3}{2}} \left\{\frac{3}{8 m^4 R^4}+\frac{1}{m^2 R^2} \left(\frac{\ln R}{8 (l+1) (2 l+1)}+\frac{12 l^4+8 l^3-33 l^2-35 l-8}{64 l (l+1)^2 (2 l+1)^2}\right) \right. \nonumber\\
&& \qquad\qquad +\left(\frac{\ln ^2R}{32 (l+1)^2 (2 l+1)^2}+\frac{\left(4 l^4-4 l^3-29 l^2-23 l-4\right) \ln R}{64 l (l+1)^3 (2 l+1)^3}\right. \nonumber\\
&& \qquad\qquad\qquad \left.\left. +\frac{12 l^4+36 l^3-l^2-58 l-29}{512 (l+1)^3 (2 l+1)^2 (2 l+3)}+C_1^-(l)\right)+\cdots \right\}, \qquad (l \neq0)
\end{eqnarray}
and
\begin{eqnarray}
\left[\varphi _2^-(m R)\right]{}_{l=0}&=&R^{\frac{3}{2}} \left\{\frac{3}{8 m^4 R^4}+\frac{1}{m^2 R^2} \left(-\frac{3 \ln R}{8}-\frac{3}{64}\right) \right. \nonumber\\
&& \qquad \left. +\left(-\frac{3 \ln ^2 R}{32}+\frac{5 \ln R}{64}-\frac{29}{1536}+C_1^-(l=0)\right)+\cdots \right\}.
\end{eqnarray}
(The initial first-derivative values also follow from these expressions, as $R$ can be regarded as a variable). Using (\ref{varphi0ip}) and (\ref{varphi0im}), we can now fix the overall constant in $\varphi_0(x)$ (see (\ref{varphi0})) also, to have
\begin{eqnarray}
\varphi _0^+(x)&=&\frac{(2 l+1) (2 l+1)!}{\sqrt{2}} \left(\frac{2}{m}\right)^{2 l+\frac{5}{2}} \sqrt{x}\; I_{2 l+2}(x), \label{varphi0p}\\
\varphi _0^-(x)&=&\frac{(2 l+1)!}{\sqrt{2}} \left(\frac{2}{m}\right)^{2 l+\frac{3}{2}} \sqrt{x}\; I_{2 l+1}(x). \label{varphi0m}
\end{eqnarray}
The consistency check here is that the small-$x$ series of (\ref{varphi0p}) and (\ref{varphi0m}), under the identification $x=m R$, should precisely match the expressions in (\ref{varphi0ip}) and (\ref{varphi0im}), respectively.

Using (\ref{recurrenceintegral}) for the first order term, we have
\begin{eqnarray}
\varphi _1(x)=\varphi _0(x) \left[\frac{\varphi _1(m R)}{\varphi _0(m R)}+\int _{m R}^x\frac{du}{\varphi _0(u){}^2} F_1(u) \right], \label{varphi1integral}
\end{eqnarray}
where we denoted
\begin{eqnarray}
F_1(x) \equiv \varphi _1'(m R) \varphi _0(m R)-\varphi _1(m R) \varphi _0'(m R) +\int _{m R}^xdt\; V_1(t) \varphi _0(t){}^2. \label{F1}
\end{eqnarray}
Here note that the original Gel'fand-Yaglom initial value problem (\ref{GYequation}) does not know about our matching point $r=R$. This then implies that, for $\varphi_1(x)$ (given by (\ref{varphi1integral})) to be related to the solution to the Galfand-Yaglom initial value problem, the right hand side of (\ref{varphi1integral}) should not depend on the specific value chosen for $R$, i.e., the $R$-dependence should disappear from the expression. As it turns out, this nontrivial consistency check is satisfied by our form (\ref{varphi1integral}). See below.

The integral in (\ref{F1}) can be performed explicitly in terms of the hypergeometric function. Since $\varphi _0(m R)$ and $\varphi _1(m R)$, given in (\ref{varphi0ip})-(\ref{varphi1im}), are in the form of the power series of $m R$ (with $m R$ taken to be small), it suffices to use the small-argument expansion for the hypergeometric function, to see that the $R$-dependences originating from the lower endpoint of the $t$-integral are completely canceled by those from $\varphi _0(m R)$ and $\varphi _1(m R)$. Therefore, $F_1(x)$ for the two sets of $J$ are given by the $R$-independent expressions
\begin{eqnarray}
F_1^+(x) &=&-\frac{(2 l+1) (2 l+3)}{4 (l+1)^2 m^3} \left(\frac{x}{m}\right)^{4 l+2}\, _2F_3 \!\left(2 l+1,2 l+\frac{5}{2};2 l+2,2 l+3,4 l+5;x^2\right), \qquad \label{F1p} \\
F_1^-(x) &=&\frac{1}{m^3} \left(\frac{x}{m}\right)^{4 l} \, _2F_3 \!\left(2 l,2 l+\frac{3}{2};2 l+1,2 l+2,4 l+3;x^2\right). \label{F1m}
\end{eqnarray}
Using these expressions for $F_1(u)$, the remaining integral in (\ref{varphi1integral}) can be performed, to obtain rather complicated expressions involving the Meijer G function \cite{meijerG}
\begin{eqnarray}
\varphi _1^+(x)&=&\varphi _0^+(x) \left\{\frac{2 l+3}{2 \sqrt{\pi }}\, G_{2,4}^{3,1}\!\left(x^2 \left|
\begin{array}{c}
 -\frac{1}{2},1 \\
 -1,0,2 l+1,-2 l-3
\end{array} \right.
\right) +\frac{2 l+3}{4 (2 l+1) ((2 l+2)!)^2} \right. \nonumber\\
&& \qquad\qquad \times\left(\frac{x}{2}\right)^{4 l+2} \frac{K_{2 l+2}(x)}{I_{2 l+2}(x)}\, _2F_3 \!\left(2 l+1,2 l+\frac{5}{2};2 l+2,2 l+3,4 l+5;x^2\right) \nonumber\\
&& \qquad \left. +\frac{\psi (2 l+1) -\ln \frac{m}{2}}{4 (l+1) (2 l+1)}-\frac{4 l^2-3}{16 (l+1)^2 (2 l+1)^2}\right\} \nonumber\\
&\equiv& -\frac{\ln m}{4 (l+1) (2 l+1)} \varphi_0^+(x) + \tilde{\varphi}_1^+ (x), \label{varphi1p} \\
\varphi _1^-(x)&=&\varphi _0^-(x) \left\{-\frac{l}{\sqrt{\pi }}\, G_{2,4}^{3,1}\!\left(x^2 \left|
\begin{array}{c}
 -\frac{1}{2},1 \\
 -1,0,2 l,-2 l-2
\end{array} \right.
\right)-\frac{1}{4 ((2 l+1)!)^2} \right. \nonumber\\
&& \qquad\qquad \times\left(\frac{x}{2}\right)^{4 l} \frac{K_{2 l+1}(x)}{I_{2 l+1}(x)} \, _2F_3 \!\left(2 l,2 l+\frac{3}{2};2 l+1,2 l+2,4 l+3;x^2\right) \nonumber\\
&& \qquad \left. -\frac{\psi (2 l+1)-\ln \frac{m}{2}}{4 (l+1) (2 l+1)}+\frac{l}{4 (l+1) (2 l+1)^2}\right\} \nonumber\\
&\equiv& \frac{\ln m}{4 (l+1) (2 l+1)} \varphi_0^-(x) + \tilde{\varphi}_1^- (x), \label{varphi1m}
\end{eqnarray}
where $\psi$ is the digamma function. These expressions are $R$-independent, the $R$ dependences from $\varphi _1(m R)/\varphi _0(m R)$ being canceled again by those from the lower end of the $u$-integral in (\ref{varphi1integral}). This nontrivial cancelation supports that our matching procedure works well. One may notice that (\ref{varphi1p}) and (\ref{varphi1m}) contain $\ln m$ dependences. These $\ln m$ terms arise after the $\ln R$ terms in (\ref{varphi1ip}) and (\ref{varphi1im}) get canceled by the $\ln (m R)$ factors originating from the low end of the $u$-integral. From the expression (\ref{varphi1p}) and (\ref{varphi1m}), we obtain the following results for $x \to \infty$:
\begin{eqnarray}
\varphi _1^+(x)&\sim& \varphi _0^+(x) \left\{\frac{-\ln \frac{m}{2}+\psi(2 l+1)}{4 (l+1) (2 l+1)}-\frac{4 l^2-3}{16 (l+1)^2 (2 l+1)^2}\right\}, \label{varphi1pasymptotic} \\
\varphi _1^-(x)&\sim& \varphi _0^-(x) \left\{\frac{\ln \frac{m}{2}-\psi(2 l+1)}{4 (l+1) (2 l+1)}+\frac{l}{4 (l+1) (2 l+1)^2}\right\}. \label{varphi1masymptotic}
\end{eqnarray}
These results will be relevant in Sec. \ref{secsmall}. (To determine the instanton effective action up to $O(m^2)$, one does not need perturbation terms higher than $\varphi_1(x)$).

We now turn to the second order term, which is given by
\begin{eqnarray}
\varphi _2(x)=\varphi _0(x) \left[\frac{\varphi _2(m R)}{\varphi _0(m R)}+\int _{m R}^x\frac{du}{\varphi _0(u){}^2} F_2(u) \right], \label{varphi2integral}
\end{eqnarray}
where we denoted
\begin{eqnarray}
F_2 (x) &\equiv& \varphi _2'(m R) \varphi _0(m R)-\varphi _2(m R) \varphi _0'(m R) \nonumber\\
&& \qquad +\int _{m R}^xdt \left\{V_1(t) \varphi _0(t) \varphi _1(t)+V_2(t) \varphi _0(t){}^2\right\}. \label{F2}
\end{eqnarray}
To find the effective action, we have to consider $\varphi_2(x)$ at $x \to \infty$. It is easy to see that the integral in (\ref{varphi2integral}) converges when $x$ approaches infinity. Thus
\begin{eqnarray}
x\to\infty \;:\; \varphi _2(x)\sim \varphi _0(x) \left[\frac{\varphi _2(m R)}{\varphi _0(m R)}+\int _{m R}^{\infty }du \frac{F_2(u)}{\varphi _0(u){}^2}\right]. \label{varphi2asymptotic}
\end{eqnarray}
Finding the second order term explicitly is not easy. However, using the small-$x$ series for $\varphi_0(x)$ and $\varphi_1(x)$, we can determine the $m R$ dependences (with $m R$ taken to be small) in the quantity inside the square bracket of (\ref{varphi2asymptotic}). This is explained below. (Note that what we need for our purpose is just the expression of $\varphi_2(x)$ for $x\to \infty$).

If small-$x$ series expansions for $\varphi_0(x)$ and $\varphi_1(x)$ (available from (\ref{varphi0p}), (\ref{varphi0m}), (\ref{varphi1p}) and (\ref{varphi1m})) are used in (\ref{F2}), we find that the $m R$ dependence in $F_2(x)$ again disappear and the result can be described in series form
\begin{eqnarray}
\frac{F_2(x)}{\varphi _0(x)^2} = \frac{D_1}{x^5} +\frac{1}{x^3} \left(E_2 \ln \frac{x}{m} +D_2\right) +\frac{1}{x} \left(E_3 \ln \frac{x}{m} +D_3\right) +\cdots \;, \label{F2series}
\end{eqnarray}
where $D_k$'s and $E_k$'s are some constants. By integrating this series, we may conclude that
\begin{eqnarray}
&& -\int _x^\infty du \frac{F_2(u)}{\varphi _0(u){}^2} \nonumber\\
&& \quad =-\frac{D_1}{4 x^4}+\frac{1}{x^2} \left(-\frac{E_2}{2} \ln \frac{x}{m}-\frac{E_2}{4}-\frac{D_2}{2}\right)+\frac{E_3}{2} \ln ^2\frac{x}{m}+D_3 \ln x+C_2+\cdots \;, \label{F2integral}
\end{eqnarray}
where $C_2$ is a certain constant. The constant $C_2$ is not given by this consideration which relies just on the small-$x$ forms of $\varphi_0(x)$ and $\varphi_1(x)$; but, if one wishes to find out its exact value, one can try the numerical evaluation based on the representation
\begin{eqnarray}
C_2&=&\lim_{x\to 0 } \left[\int _x^\infty du \frac{F_2(u)}{\varphi _0(u){}^2} \right. \nonumber\\
&& \qquad \left. -\frac{D_1}{4 x^4}+\frac{1}{x^2} \left(-\frac{E_2}{2} \ln \frac{x}{m}-\frac{E_2}{4}-\frac{D_2}{2}\right)+\frac{E_3}{2} \ln ^2\frac{x}{m}+D_3 \ln x\right]. \label{C2def}
\end{eqnarray}
Explicit evaluation of the constant $C_2$ is given in the Appendix. Using (\ref{F2integral}) in (\ref{varphi2asymptotic}), under the identification $x=m R$, we find that all $R$ dependences cancel again: so we are assured that our matching procedure works fine for the second-order term also. By this procedure we are led to the following expressions as $x$ is taken to be very large:
\begin{eqnarray}
\varphi _2^+(x) &\sim& \varphi _0^+(x) \left\{\frac{\ln ^2 m}{32 (l+1)^2 (2 l+1)^2}-\frac{(3 l+2) \left(4 l^2+4 l-1\right) \ln m}{32 (l+1)^3 (2 l+1)^3 (2 l+3)} \right. \nonumber\\
&& \qquad \left. -\frac{(4 l+5) \left(4 l^3+20 l^2+27 l+7\right)}{64 (l+1) (l+2) (2 l+1)^3 (2 l+3)^2}+C_1^+(l)+C_2^+(l)\right\},  \label{varphi2pasymptotic} \\
\varphi _2^-(x)&\sim& \varphi _0^-(x) \left\{\frac{\ln ^2m}{32 (l+1)^2 (2 l+1)^2}+\frac{\left(6 l^3+17 l^2+12 l+2\right) \ln m}{32 l (l+1)^3 (2 l+1)^3} \right. \nonumber\\
&& \qquad \left. +\frac{8 l^3+22 l^2+25 l+12}{256 l (l+1)^3 (2 l+1) (2 l+3)}+C_1^-(l)+C_2^-(l)\right\}, \qquad (l \neq 0)
\end{eqnarray}
and
\begin{eqnarray}
\left[\varphi _2^-(x)\right]_{l=0}\sim \left[\varphi _0^-(x)\right]_{l=0} \left\{-\frac{3 \ln ^2 m}{32} -\frac{\ln m}{8}-\frac{7}{768}+C_1^-(l=0)+C_2^-(l=0)\right\}. \label{varphi2masymptotic}
\end{eqnarray}
The constants $C_2^\pm (l)$ here may be found using the representation in (\ref{C2def}). [But, unlike $C_1^\pm (l)$, $C_2^\pm (l)$ have some $\ln m$ dependent pieces; see (\ref{C2plog})-(\ref{C2m0log}) below]. In the next section these results will be used to find the instanton effective action up to $O(m^4)$.

\section{Small mass behaviors of the instanton determinant} \label{secsmall}
In the previous section we derived the asymptotic expressions that the solutions to the Gel'fand-Yaglom initial value problems when mass is small. In this section we will use those results to determine small-mass behavior of the instanton effective action. Using the Gel'fand-Yaglom formula, the sum of the partial wave contributions $\Gamma_{(l,j=l+\frac{1}{2})}(m)$ and $\Gamma_{(l+\frac{1}{2},j=l)}(m)$ (defined in (\ref{partialeffectiveaction})) can be expressed as
\begin{eqnarray}
&& \Gamma_{(l,j=l+\frac{1}{2})}(m)+\Gamma_{(l+\frac{1}{2},j=l)}(m) \nonumber\\
&& =\lim_{x\to \infty }  \ln \!\left(\frac{[\varphi_0 ^+(x)+m^2 \varphi_1 ^+(x)+m^4 \varphi_2 ^+(x)+\cdots] [\varphi_0 ^-(x)+m^2 \varphi_1 ^-(x)+m^4 \varphi_2 ^-(x)+\cdots]}{\varphi ^{\text{free}}(x) \left[\varphi ^{\text{free}}(x)\right]_{l=l+\frac{1}{2}} }\right) \nonumber\\
&& \equiv \Gamma_l^{(0)} +m^2 \Gamma_l^{(1)} +m^4 \Gamma_l^{(2)} +O(m^6). \label{partialdet}
\end{eqnarray}
Then, using (\ref{varphi0p}), (\ref{varphi0m}), (\ref{varphi1pasymptotic}), (\ref{varphi1masymptotic}) and (\ref{varphi2pasymptotic})-(\ref{varphi2masymptotic}), we obtain following results:
\begin{eqnarray}
\Gamma^{(0)}_l &=& \ln \!\left(\frac{2 l+1}{2 l+2}\right), \label{partialdetm0} \\
\Gamma^{(1)}_l &=& \frac{(4 l+3)}{16 (l+1)^2 (2 l+1)^2}, \label{partialdetm2} \\
\Gamma^{(2)}_l &=& \left\{\frac{\psi(2 l+1)+\ln 2}{8 (l+1)^2 (2 l+1)^2}-\frac{16 l^4-32 l^3-134 l^2-93 l-12}{64 l (l+1)^3 (2 l+1)^3 (2 l+3)}\right\} \ln m \nonumber\\
&& -\frac{(\psi(2 l+1)+\ln 2)^2}{16 (l+1)^2 (2 l+1)^2}+\frac{\left(8 l^2+4 l-3\right) (\psi (2 l+1)+\ln 2)}{64 (l+1)^3 (2 l+1)^3} \nonumber\\
&& -\frac{384 l^8+2240 l^7+4560 l^6+2776 l^5-2992 l^4-5702 l^3-3529 l^2-1034 l-144}{512 l (l+1)^4 (l+2) (2 l+1)^4 (2 l+3)^2} \nonumber\\
&& +C_1^+(l) +C_1^-(l) +C_2^+(l) +C_2^-(l), \qquad (l \neq 0) \label{partialdetm4}
\end{eqnarray}
and
\begin{eqnarray}
\Gamma_{l=0}^{(2)} &=&-\frac{\ln^2 m}{8} +\frac{\ln m}{8} \left(\ln 2-\gamma -\frac{11}{24}\right) -\frac{263}{4608}-\frac{(\ln 2-\gamma )^2}{16} -\frac{3(\ln 2-\gamma )}{64} \nonumber\\
&& \qquad +C_1^+(l=0) +C_1^-(l=0) +C_2^+(l=0) +C_2^-(l=0), \label{partialdetm40}
\end{eqnarray}
where $\gamma=0.577216\ldots$ is Euler's constant.

For the leading term, i.e., $\Gamma^{(0)}_l$, which is appropriate to the zero mass limit, our result (\ref{partialdetm0}) was also obtained in \cite{insdet}. Using the result (\ref{partialdetm0}) for $\Gamma_l$ (and (\ref{largeL}) with $m=0$) into the formula (\ref{effectiveaction}) yields the zero mass limit of the effective action
\begin{eqnarray}
\tilde{\Gamma}^S(m=0)&=&\lim_{L\to \infty } \left[\sum _{l=0,\frac{1}{2},\cdots }^L (2 l+1) (2 l+2) \ln \!\left(\frac{2 l+1}{2 l+2}\right)+2 L^2+4 L-\frac{\ln L}{6} +\frac{127}{72}-\frac{\ln 2}{3} \right] \nonumber\\
&=&\alpha\! \left(\frac{1}{2}\right)=0.145873\ldots \;.
\end{eqnarray}
This is precisely the 't Hooft value \cite{thooft}. By using the result (\ref{partialdetm2}) in (\ref{effectiveaction}) (together with (\ref{largeL})), we can find the $O(m^2)$ contribution to the effective action. Explicitly, this $O(m^2)$ contribution to $\tilde{\Gamma}^S(m)$ is
\begin{eqnarray}
&& \lim_{L\to \infty } m^2 \left[\sum _{l=0,\frac{1}{2},\cdots }^L \frac{4 l+3}{8 (l+1) (2 l+1)}+\frac{1}{2} \left( -\ln L+\ln \frac{m}{4} +1 \right) \right] \nonumber\\
&& = \frac{m^2}{2} \left\{\ln \frac{m}{2}+\gamma +\frac{1}{2}\right\}. \label{m2coef}
\end{eqnarray}
While the $m^2 \ln m$ term of this expression agrees with the previous result \cite{calitz}, the numerical coefficient of the $m^2$ term is different from the value given in \cite{kwon}. We believe that (\ref{m2coef}) is correct, since we no longer think that the argument used in Appendix B of Ref. \cite{kwon}, which is crucial to obtain the old value, is justified.

If we use also the result (\ref{partialdetm4}) and (\ref{partialdetm40}) in (\ref{effectiveaction}), we can obtain the $O(m^4)$ contribution to the effective action. Here we remark that $C_2(l)$ contains some $\ln m$ dependences, while $C_1(l)$ is $m$-independent. The $\ln m$ terms in $C_2(l)$ originate from our expression for $\varphi_1(x)$, given in (\ref{varphi1p}) and (\ref{varphi1m}). Since the $\ln m$ terms in $\varphi_1(x)$ are proportional to $\varphi_0(x)$, the integral in (\ref{C2def}) for the $\ln m$ terms is essentially the same as the integral we have in (\ref{varphi1integral}). Thus the $\ln m$ dependences in $C_2(l)$ can be found explicitly and so we write
\begin{eqnarray}
C_2^+(l)&=&\left\{\frac{4 l^3+2 l^2-10 l-7}{32 (l+1)^3 (2 l+1)^3 (2 l+3)}-\frac{\psi(2 l+1)+\ln 2}{16 (l+1)^2 (2 l+1)^2}\right\} \ln m + \tilde{C}_2^+(l), \label{C2plog} \\
C_2^-(l)&=&\left\{\frac{4 l^2-3}{64 (l+1)^3 (2 l+1)^3}-\frac{\psi(2 l+1)+\ln 2}{16 (l+1)^2 (2 l+1)^2}\right\} \ln m + \tilde{C}_2^-(l), \quad (l \neq 0) \label{C2mlog}
\end{eqnarray}
and
\begin{eqnarray}
C_2^-(l=0) = \frac{3 \ln m}{16} \left(\frac{3}{4}-\gamma +\ln 2\right)+ \tilde{C}_2^-(l=0). \label{C2m0log}
\end{eqnarray}
Here, $m$-independent constants $\tilde{C}_2(l)$ have the representation
\begin{eqnarray}
\tilde{C}_2(l)&=&\lim_{x\to 0 } \left[\int _x^\infty du \frac{\tilde{F}_2(u)}{\varphi _0(u){}^2} \right. \nonumber\\
&& \qquad \left. -\frac{D_1}{4 x^4}+\frac{1}{x^2} \left(-\frac{E_2}{2} \ln x-\frac{E_2}{4}-\frac{D_2}{2}\right)+\frac{E_3}{2} \ln ^2x+D_3 \ln x\right], \label{tildeC2def}
\end{eqnarray}
where $\tilde{F}_2(x)$ is defined by (\ref{F2}) with the replacement $\varphi_1(x) \to \tilde{\varphi}_1(x)$ (see (\ref{varphi1p}) and (\ref{varphi1m}) for $\tilde{\varphi}_1(x)$), and $D_k$'s and $E_k$'s are the same constants that appear in (\ref{F2series}). Using (\ref{partialdetm4}) and (\ref{partialdetm40}) together with (\ref{C2plog})-(\ref{C2m0log}) in (\ref{effectiveaction}), we can obtain the $O(m^4)$ contribution to the effective action. This $O(m^4)$ contribution to $\tilde{\Gamma}^S(m)$ is
\begin{eqnarray}
m^4 \left\{ -\frac{\ln^2 m}{4}+\frac{\ln m}{2} \left(\frac{1}{2}-\gamma +\ln 2\right)+C \right\},
\end{eqnarray}
where the constant $C$, as derived from (\ref{partialdetm4}) and (\ref{partialdetm40}), has the value
\begin{eqnarray}
&& C = \sum_{l=0,\frac{1}{2},\cdots}^\infty \left[-\frac{(\psi(2 l+1)+\ln 2)^2}{8 (l+1) (2 l+1)}+\frac{\left(8 l^2+4 l-3\right) (\psi(2 l+1)+\ln 2)}{32 (l+1)^2 (2 l+1)^2} \right. \nonumber\\
&& \qquad +(2 l+1) (2 l+2) \left\{ C_1^+(l) +C_1^-(l) +\tilde{C}_2^+(l) +\tilde{C}_2^-(l) \right\} \Bigg]-\frac{263}{2304} \nonumber\\
&& -\!\!\sum_{l=\frac{1}{2},1,\cdots}^\infty \!\!\frac{ 384 l^8+2240 l^7+4560 l^6+2776 l^5-2992 l^4-5702 l^3-3529 l^2 -1034 l-144 } {256 l (l+1)^3 (l+2) (2 l+1)^3 (2 l+3)^2 }. \qquad\;\; \label{Cdef}
\end{eqnarray}

If we put together our findings, the scalar effective action $\tilde{\Gamma}^S(m \rho)$ for small $m \rho$ (we restored the instanton size parameter $\rho$ here) can be expressed by the formula
\begin{eqnarray}
\tilde{\Gamma}^S (m \rho)&=&\alpha\! \left(\frac{1}{2}\right)+\frac{(m \rho)^2}{2} \left\{\ln \frac{m \rho}{2}+\gamma +\frac{1}{2}\right\} \nonumber\\
&& +(m \rho)^4 \left\{-\frac{\ln^2 (m \rho)}{4}+\frac{\ln (m \rho)}{2} \left(\frac{1}{2}-\gamma +\ln 2\right)+C\right\}+O\! \left((m \rho)^6\right). \label{finalresult}
\end{eqnarray}
Then, using this expression with our formulas (\ref{gammaS}) and (\ref{gammaF}), we obtain the small-mass behavior of the fermion effective action as given in the abstract. To evaluate the constant $C$, a rather laborious numerical routine is required. (See the Appendix). With some effort, we determined this value to be
\begin{eqnarray}
C=-0.382727\ldots \;.
\end{eqnarray}

\section{Discussion} \label{secdiscussion}
\begin{figure}
\includegraphics[scale=1.7]{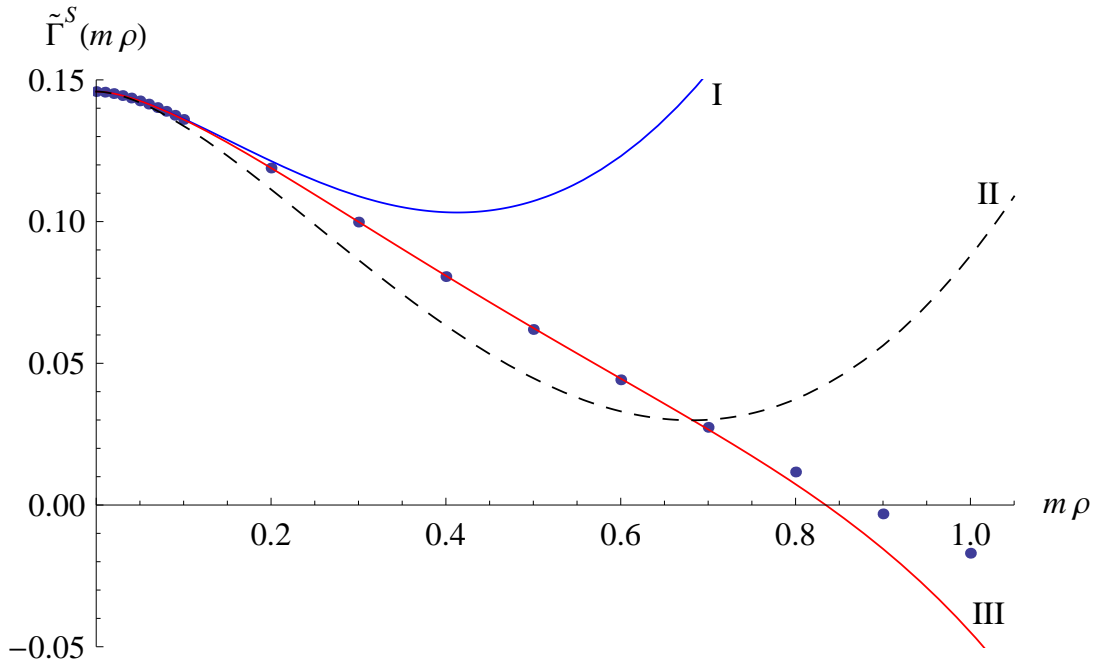}\\
\caption{The result of the small mass expansion for $\tilde{\Gamma}^S(m \rho)$, curves I and III, against the exact numerical data (dots) from \cite{insdet}. The curve III is based on our formula (\ref{finalresult}) considered up to $O((m \rho)^4)$, while the curve I contains the expansion terms only up to $O((m \rho)^2)$. The curve II is obtained if one uses the wrong formula from \cite{kwon}.} \label{fig1}
\end{figure}
In Fig. \ref{fig1} we present the predictions based on our formula (\ref{finalresult}) against the (essentially exact) numerical result for the same quantity first provided in Ref. \cite{insdet}. Notice that the curve I --- the predictions based on our formula (\ref{finalresult}) considered only up to $O((m \rho)^2)$ terms --- matches the numerical result (represented by dots in the figure) closely for $m \rho \lesssim 0.2$, but not so for $m \rho$ larger than $0.2$. [The curve II, which is based on the wrong formula suggested in Ref. \cite{kwon}, starts to deviate from the numerical result at much smaller value of $m \rho$]. On the other hand, the curve III --- the predictions based on our formula (\ref{finalresult}) now including up to $O((m \rho)^4)$ terms --- can reproduce quite faithfully numerical results all the way up to $m \rho \approx 0.8$.
\begin{figure}
\includegraphics[scale=1.7]{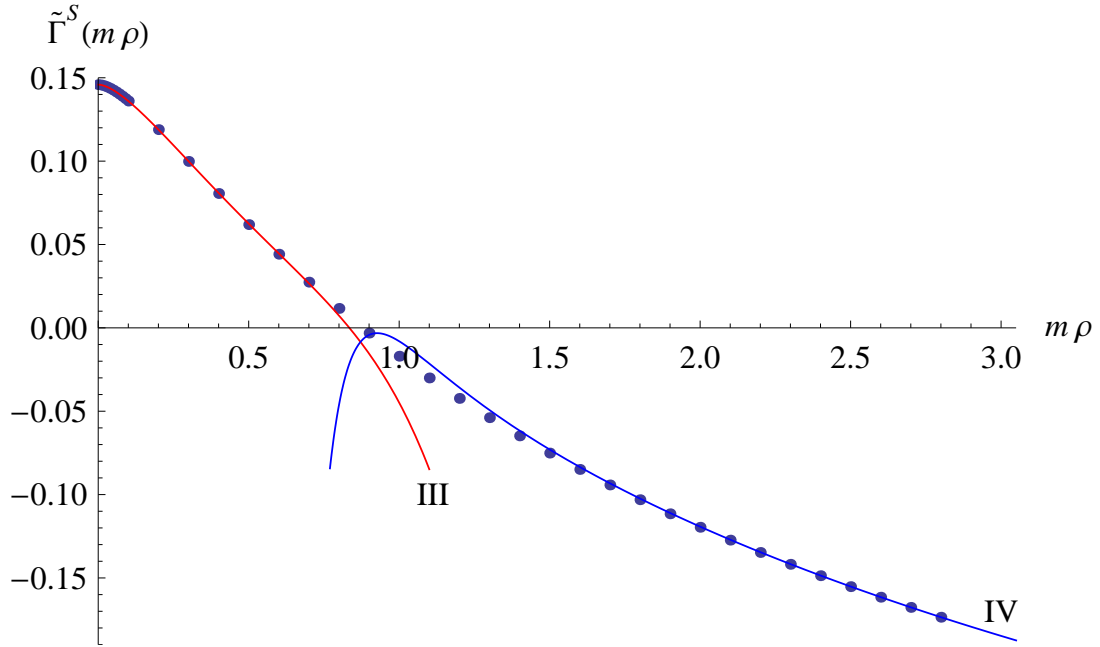}\\
\caption{Plot of our small mass expansion (\ref{finalresult}) (the curve III) and the large mass expansion from \cite{kwon} (the curve IV), together with the exact numerical result (dots) for $\tilde{\Gamma}^S(m \rho)$.} \label{fig2}
\end{figure}
In Fig. \ref{fig2} we have shown this curve III together with the predictions due to the large-mass expansion (the curve IV), which were obtained earlier in Ref. \cite{kwon}. As one can see, entire numerical results (represented by dots) can be reconstructed rather faithfully just by connecting the small-mass-expansion-based curve with the large-mass-expansion-based curve at the mid-point (close to the value $m \rho =1$) where the two curves meet!

\begin{table}
\caption{\label{table1}Comparision of the small mass expansion and the exact values}
\begin{ruledtabular}
\begin{tabular}{ccccc}
$m\rho$ & up to $O((m\rho)^2)$ &up to $O((m\rho)^4)$ & Exact value & Error  \\
\hline
0.01	&0.145662	&0.145662	&0.14566222   &$<10^{-6}$ \\
0.04	&0.143605	&0.143595	&0.14359532   &$<10^{-6}$ \\
0.07	&0.140299   &0.140227	&0.14022767   &$<10^{-6}$ \\
0.1		&0.136281	&0.136039	&0.13603823	  &0.000001 \\
0.2		&0.121366	&0.118924	&0.11890459   &0.000019 \\
0.3		&0.108978	&0.0999389	&0.09982892   &0.000056 \\
0.4		&0.103296	&0.0809004	&0.08059098   &0.000308 \\
0.5		&0.107238	&0.0624694	&0.06193515   &0.000534 \\
0.6		&0.123057	&0.0446128	&0.04417211   &0.000441 \\
0.7		&0.152585	&0.0266823	&0.02740940   &$-0.000727$
\end{tabular}
\end{ruledtabular}
\end{table}

For more explicit comparison, numerical values of the small mass expansion and exact data for several $m\rho$ values are presented in Table~\ref{table1}. Two cases of the small mass expansion (one with the terms only up to $O((m \rho)^2)$ and the other up to $O((m \rho)^4)$) are tabulated.  In the first case,  the values are close to exact values only when $m\rho<0.1$.  Note that, in the second case, the errors between  approximate values and exact values are less than $ 10^{-6}$ when $m\rho<0.1$. The errors increase as $m\rho$ increases but it is still less than $7\times10^{-4}$ even when $m\rho=0.7$.

Finally we note that Fucci and Kirsten recently have developed a similar small mass expansion in a different situation when the background space is a generalized cone \cite{kirsten}.

\section*{Acknowledgments}
This work was supported in part by Basic Science Research Program through the National Research Foundation of Korea (NRF) funded by the Ministry of Education, Science and Technology: 2009-0076297(CL) and KRF-2008-313-C00175(HM).

\newpage
\appendix
\section{Numerical procedures to find the constant $C$}
In this Appendix we will describe the numerical method used to find the constant $C$ (see (\ref{Cdef})). First we have to determine the values of $C_1^\pm (l)$ and $\tilde{C}_2^\pm (l)$. For $C_1^\pm (l)$, we utilize (\ref{C1def}) and perform the integral given there explicitly for each $l$ value. When $2 l$ is an integer, $\phi _1(r)$ (given in (\ref{phi1plussol}) and (\ref{phi1minussol})) can actually be expressed in simpler form. For instance, for $l=0$,
\begin{eqnarray}
\left[\phi_1^+(r)\right]_{l=0} &=& \frac{\left(r^2+1\right)^{\frac{3}{2}} \ln \!\left(r^2+1\right)}{16 \sqrt{r}}+\frac{r^{\frac{3}{2}} \left(4 r^4+3 r^2-6\right)}{96 \sqrt{r^2+1}}, \\
\left[\phi_1^-(r)\right]_{l=0} &=& -\frac{\left(r^2+1\right)^{\frac{3}{2}} \ln \!\left(r^2+1\right)}{8 r^{\frac{3}{2}}}+\frac{\sqrt{r} \left(2 r^4+3 r^2+2\right)}{16 \sqrt{r^2+1}}
\end{eqnarray}
and $\phi _1^\pm(r)$ for other $l$ values can be written in similar simple forms as well. With these expressions, the integral in (\ref{C1def}) can be performed in closed form. For instance, for $l=0$, we have
\begin{eqnarray}
&& \int _0^r\frac{du}{\phi_0^+(u){}^2} \int _0^udt\; \phi_0^+(t) \phi_1^+(t) \nonumber\\
&& \quad = -\frac{\text{Li}_2\left(-r^2\right)}{64}+\frac{\left(r^2+1\right) \left(2 r^4-5\right) \ln \!\left(r^2+1\right)}{192 r^2 \left(r^2+2\right)}+\frac{3 r^6-14 r^4-21 r^2+30}{1152 \left(r^2+2\right)}, \\
&& \int _0^r\frac{du}{\phi_0^-(u){}^2} \int _0^udt\; \phi_0^-(t) \phi_1^-(t) \nonumber\\
&& \quad= \frac{3 \text{Li}_2\left(-r^2\right)}{64}+\frac{2 r^6+16 r^4-21 r^2-6}{384 r^2}+\frac{\left(-r^6+2 r^4+4 r^2+1\right) \ln \!\left(r^2+1\right)}{64 r^4},
\end{eqnarray}
where $\text{Li}_2(-r^2)$ is the polylogarithm function. Analogous calculation can be done for $l=\frac{1}{2},1,\frac{3}{2},\cdots$ also. Using these expressions with (\ref{C1def}), we can determine $C_1^\pm (l)$ explicitly for each $l$-value, to find the results like
\begin{eqnarray}
&& C_1^+ (0) = \frac{31}{1152}+\frac{\pi ^2}{384},\quad C_1^+ \!\left(\frac{1}{2}\right) = \frac{419}{207360}+\frac{\pi ^2}{3456},\; \cdots \;, \\
&& C_1^- (0) = -\frac{9}{128}-\frac{\pi ^2}{128},\quad C_1^- \!\left(\frac{1}{2}\right) = \frac{145}{10368}+\frac{\pi ^2}{3456},\; \cdots \;.
\end{eqnarray}

To find $\tilde{C}_2^\pm (l)$, we use (\ref{tildeC2def}): but, here, the integral cannot be performed explicitly. Therefore, we resort to a numerical method. Since the integral in (\ref{tildeC2def}) diverges when $x$ approaches zero, it is better to divide the integration range into two parts. With an appropriate choice of the intermediate value $u=u^*$, the small-$x$ series expansion will be a good approximation for $u \leq u^*$ (i.e., use the series (\ref{F2series}) for $F_2(u)$, including higher order terms), and for $u \geq u^*$ the integral can be performed numerically to obtain a finite value. By this method, numerical value for $\tilde{C}_2^\pm (l)$ can be obtained with some effort. The values for $l=0$ and $l=\frac{1}{2}$, for instance, are given as
\begin{eqnarray}
&& \tilde{C}_2^+ (0) = 0.0517859\ldots\;,\quad \tilde{C}_2^+ \!\left(\frac{1}{2}\right) =0.0145914\ldots\;, \\
&& \tilde{C}_2^- (0) = -0.0505132\ldots\;,\quad \tilde{C}_2^- \!\left(\frac{1}{2}\right) = -0.0294879\ldots\;.
\end{eqnarray}

Using the thus determined values of $C_1^\pm(l)$ and $\tilde{C}_2^\pm (l)$ with our formula (\ref{Cdef}), we can determine $C$ numerically. But, since the $l$-sum converges rather slowly, it is desirable to employ some acceleration technique. Here, to achieve faster convergence for the $l$-sum, we may utilize the expressions of higher-order WKB terms (up to $O((\frac{1}{L})^6)$) for the quantity $\Gamma_{l>L}(m)$  (see Refs. \cite{rea1,rea3} for detailed discussions). That is, after performing the $l$-sum in (\ref{Cdef}) up to $l=L$, we add a correction term which can be read off from the coefficient of $m^4$ for the quantity $\Gamma_{l>L}(m)$,
\begin{eqnarray}
&& \frac{1}{L^2} \left\{-\frac{3}{16} \ln \!\left(4 L\right)+\frac{5}{32}\right\}+\frac{1}{L^3} \left\{\frac{3}{8} \ln \!\left(4 L\right)-\frac{1}{2}\right\}+\frac{1}{L^4} \left\{-\frac{19}{32} \ln \!\left(4 L\right)+\frac{1021}{1024}\right\} \nonumber\\
&& \quad +\frac{1}{L^5} \left\{\frac{7}{8} \ln \!\left(4 L\right)-\frac{437}{256}\right\}+\frac{1}{L^6} \left\{-\frac{161}{128} \ln \!\left(4 L\right)+\frac{168517}{61440}\right\} +\cdots \;.
\end{eqnarray}
With this correction term, the convergence of the $l$-sum becomes very fast: we can take $L=10$ to obtain the accurate numerical result $C=-0.382727\ldots$.

\end{document}